\documentclass[aps,twocolumn,showkeys,showpacs,preprintnumbers,prd,superscriptaddress,nofootinbib,10pt]{revtex4-1}
\usepackage{graphicx,epsf,bm,amsmath,amsfonts,amssymb,epstopdf,natbib,hyperref,color,verbatim,multirow,bm,lmodern,tensor,appendix}
\hypersetup{colorlinks=true,urlcolor=blue,citecolor=blue,linkcolor=blue,menucolor=blue,anchorcolor=blue,filecolor=blue}

\date{\today}

\begin{document}

\title{Importance of being nonminimally coupled:\\Scalar Hawking radiation from regular black holes}

\author{Marco Calz\`{a}}
\email{marco.calza@unitn.it}
\affiliation{Department of Physics, University of Trento, Via Sommarive 14, 38123 Povo (TN), Italy}
\affiliation{Trento Institute for Fundamental Physics and Applications (TIFPA)-INFN, Via Sommarive 14, 38123 Povo (TN), Italy}

\author{Massimiliano Rinaldi}
\email{massimiliano.rinaldi@unitn.it}
\affiliation{Department of Physics, University of Trento, Via Sommarive 14, 38123 Povo (TN), Italy}
\affiliation{Trento Institute for Fundamental Physics and Applications (TIFPA)-INFN, Via Sommarive 14, 38123 Povo (TN), Italy}

\author{Sunny Vagnozzi}
\email{sunny.vagnozzi@unitn.it}
\affiliation{Department of Physics, University of Trento, Via Sommarive 14, 38123 Povo (TN), Italy}
\affiliation{Trento Institute for Fundamental Physics and Applications (TIFPA)-INFN, Via Sommarive 14, 38123 Povo (TN), Italy}

\date{\today}

\begin{abstract}
\noindent In curved space-time, a scalar field $\phi$ is generically expected to couple to curvature, via a coupling of the form $\xi\phi^2R$. Yet in the study of Hawking emission from regular black holes (RBHs), where scalar fields are often introduced as test field probes of the geometry, and the Ricci scalar is generically non-zero, this non-minimal coupling is almost always ignored. We revisit this assumption by studying scalar Hawking emission from four representative RBHs (the Bardeen, Hayward, Simpson-Visser, and D'Ambrosio-Rovelli space-times), within two benchmark cases: the conformal case $\xi=1/6$, and a large negative value $\xi=-10^4$ motivated by Higgs inflation. We compute the graybody factors and emission spectra, showing that the latter can be either enhanced or suppressed, even by several orders of magnitude. A crucial role is played by the sign of the term $\xi fR$, with $f(r)=-g_{tt}$ in Schwarzschild-like coordinates, as it determines whether the non-minimal coupling suppresses or enhances the geometric potential barrier. For the D'Ambrosio-Rovelli case with large negative $\xi$, the low-energy emission spectrum is enhanced by up to five orders of magnitude, since $\xi fR<0$ throughout the space-time, leading to a deep potential well which broadens the transmissive window. The deviations we find can be particularly relevant in the case where primordial RBHs are dark matter candidates, given the impact of the non-minimal coupling on their evaporation history.
\end{abstract}

\maketitle
\section{Introduction}

Scalar fields appear ubiquitously in a wide range of areas in physics, partly because of their simplicity and ability to capture the dynamics of more complex systems while remaining tractable. In theoretical high-energy physics and cosmology, (pseudo)scalar fields play a central role, appearing in contexts ranging from the Higgs mechanism~\cite{Englert:1964et,Higgs:1964ia,Guralnik:1964eu} to the simplest models of cosmic inflation~\cite{Albrecht:1982wi,Linde:1983gd,Freese:1990rb,Martin:2013tda,Nojiri:2014zqa,Myrzakulov:2015qaa,Ema:2017rqn,Oikonomou:2022bqb,Oikonomou:2022tux,Odintsov:2023weg}, dark matter~\cite{Hu:2000ke,Boehm:2003hm,Cline:2013gha,Foot:2014uba,Marsh:2015xka,Hui:2016ltb,Odintsov:2019mlf,Odintsov:2019evb,Ferreira:2020fam,Oikonomou:2023kqi}, and dark energy~\cite{Ratra:1987rm,Wetterich:1987fm,Caldwell:1997ii,Carroll:1998zi,Zlatev:1998tr,Amendola:1999er,Khoury:2003rn,Hinterbichler:2010es,Odintsov:2020nwm,Oikonomou:2020qah}, while string compactifications are generically expected to lead to the existence of hundreds of such particles (the so-called ``string axiverse'')~\cite{Svrcek:2006yi,Arvanitaki:2009fg,Cicoli:2012sz,Visinelli:2018utg,Cicoli:2023opf}; on the other hand, effective scalar degrees of freedom emerge, for instance, in modified gravity theories~\cite{Sotiriou:2008rp,DeFelice:2010aj,Nojiri:2010wj,Chamseddine:2013kea,Cai:2015emx,Nojiri:2017ncd,Burrage:2017qrf,Oikonomou:2022wuk,Hell:2023mph,Hell:2025wha,Heisenberg:2025fxc,Hell:2025lbl,Barker:2025gon} and in the low-energy description of QCD~\cite{Pelaez:2015qba}. Moreover, the search for new (typically light) scalar fields is a major focus of present-day experimental high-energy physics, astroparticle physics, and cosmology, with numerous dedicated efforts now underway, that already probe or exclude key theoretical benchmarks~\cite{ADMX:2010ubl,Arvanitaki:2014wva,Brito:2015oca,SimonsObservatory:2018koc,Vagnozzi:2021quy,Odintsov:2022cbm,Poulin:2023lkg,Alesini:2023qed,Tsai:2023zza,Feleppa:2025clx}. In condensed matter physics, scalar order parameters appear in the description of phenomena such as phase transitions, superconductivity, superfluidity, and Bose-Einstein condensation within the Landau-Ginzburg framework~\cite{Hohenberg:1977ym}. Finally, scalar fields play an important role in gravitational physics, particularly when it comes to probing the effects of space-time curvature in the strong-gravity regime, for instance around black holes (BHs).

A generic expectation within the context of quantum field theories in curved space-times is that scalar fields should be non-minimally coupled to the space-time curvature. This \textit{non-minimal coupling} requirement becomes especially relevant when the curvature is large~\cite{Birrell:1982ix,Parker:2009uva,Fulling:1989nb}. To be more concrete, let us consider the following action, in the framework of General Relativity (GR), for a real scalar field $\phi$:~\footnote{Note that we use the mostly plus signature for the metric $g_{\mu\nu}$.}
\begin{equation}
\begin{aligned}
S = \int d^4x\sqrt{g}\, \Bigg[ &\frac{M_{\text{Pl}}^2}{2}R-\frac{1}{2} g^{\mu\nu} \partial_\mu \phi \partial_\nu \phi - \frac{\xi}{2} \phi^2 R \\
&- \frac{1}{2} m^2 \phi^2 - V(\phi) \Bigg] \,,
\label{eq:action}
\end{aligned}
\end{equation}
where $\sqrt{g} \equiv \sqrt{-\det(g_{\mu\nu})}$, $M_{\text{Pl}}$ is the reduced Planck mass, $\xi$ is a dimensionless parameter controlling the strength of the non-minimal coupling, whereas $m$ and $V(\phi)$ are respectively the scalar field mass and potential. The $\xi=0$ case corresponds to a minimally coupled scalar field. A special case is that of a massless, free scalar field (i.e.\ $m=V=0$) with non-minimal coupling strength $\xi=1/6$. In this particular case, the action of Eq.~(\ref{eq:action}) turns out to be conformally invariant. That is, assuming that $g_{\mu\nu}(x)$ is a solution to the equations of motion, then any other conformally transformed metric $\bar{g}_{\mu\nu}(x)=\Omega^2(x)g_{\mu\nu}(x)$ will also be a solution, for any smooth function $\Omega(x)$. This also implies that conformal rescaling preserves the form of the Klein-Gordon (KG) equation, given by the following:
\begin{equation}
\square \phi -\xi\phi R=0\,.
\label{eq:kgm0}
\end{equation}
It has also been shown that the two-point function of a non-minimally coupled massive scalar field reduces to the Minkowski limit if, and only if, $\xi=1/6$, thus extending the importance of this value also to theories which are not conformally invariant~\cite{Faraoni:1991xe,Sonego:1993fw}. Importantly, even leaving aside considerations on conformal symmetry, renormalization of interacting scalar field theories in curved space-time inevitably requires the addition of a counterterm proportional to $\phi^2R$ (see e.g.\ the case $\lambda\phi^4$ case~\cite{Bunch:1980br}). Therefore, even if absent at tree level, a non-minimal coupling of the form given in Eq.~(\ref{eq:action}) will appear at loop level, and can be considered a generic feature of scalar fields in curved space-time, with stability arguments requiring that $\xi\leq 0$ or $\xi\geq 1/6$~\cite{Hosotani:1985at}.

The non-minimal coupling can play a very important role also in cosmological settings, for instance within cosmic inflation~\cite{Faraoni:1996rf,Faraoni:2000wk}. An important example is that of Higgs inflation, where inflation is driven by the non-minimal coupling of the Standard Model Higgs field to curvature~\cite{Bezrukov:2007ep}. For the model to be consistent with cosmological observations, a very large non-minimal coupling is required ($\vert \xi \vert \gtrsim 10^4$), which in itself raises questions about naturalness and unitarity violation~\cite{Barbon:2009ya,Burgess:2010zq,Bezrukov:2010jz,Giudice:2010ka,Calmet:2013hia,Rubio:2015zia,George:2015nza,Fumagalli:2016lls,Fumagalli:2017cdo,Rubio:2018ogq} (see also Refs.~\cite{Herranen:2014cua,Herranen:2015ima,Salvio:2015kka}). This problem can be alleviated in some scale-invariant models of inflation, where the required value of the non-minimal coupling strength can be drastically reduced~\cite{Rinaldi:2015uvu,Tambalo:2016eqr,Vicentini:2019etr,Ghoshal:2022qxk,Cecchini:2024xoq}. It is interesting to note that negative values of $\xi$ ensure that the effective Planck mass $(M_{\text{Pl}}^2-\xi\phi^2)/2$ is positive when $\phi\sim M_{\text{Pl}}$. Still in the context of early Universe physics, the time-evolution of the quantized matter fields and the associated stress tensors is crucial as it can populate an expanding and empty universe with particles, provided the associated fields are not conformally coupled to curvature~\cite{Parker:1969au,Parker:1971pt}. Such a mechanism, called preheating, is often invoked when the universe exits from inflation and needs to be filled with ordinary matter particles~\cite{Bassett:2005xm}, and can be strongly affected by non-minimal couplings, which can drastically enhance the particle production process~\cite{DeCross:2015uza,DeCross:2016fdz,DeCross:2016cbs}.

As alluded to earlier, scalar fields play an extremely important role also in the strong-gravity regime, serving as simple testbeds for exploring the dynamics of BHs, and how these objects respond to perturbations. For instance, such fields are widely employed to investigate spectral characteristics of BHs such as quasinormal modes and graybody factors (GBFs), as well as phenomena such as superradiant instabilities and the possible formation of long-lived scalar clouds around BHs. For scalar fields on the simplest BH backgrounds such as the Schwarzschild and Kerr ones, the KG equation does not depend on the non-minimal $\xi$ since the Ricci scalar $R$ is zero and the corresponding counterterm vanishes, at least at 1-loop. From the classical point of view, the analysis of the KG equation is crucial to calculate the GBFs and spectra of emitted particles. From a quantum point of view, finding solution modes is the first step to quantize the field, and calculate $n$-point correlation functions~\cite{Fontana:2023zqz} and quantum stress tensors~\cite{Fabbri:2005mw}.

Nevertheless, both the Schwarzschild and Kerr space-times are plagued by pathological singularities, highlighting one of the most pressing problems of classical GR, captured by the Penrose-Hawking singularity theorems. This provides strong motivation for the study of so-called \textit{regular} BH space-times, which are free of singularities and may provide a more consistent stage for quantum field theory in curved space-time (see e.g.\ Refs.~\cite{Sebastiani:2022wbz,Lan:2023cvz,Carballo-Rubio:2025fnc} for reviews). Significant effort has been devoted to the study of regular BHs, starting from the pioneering works of Bardeen~\cite{Bardeen:1968ghw} and Hayward~\cite{Hayward:2005gi}, see e.g.\ Refs.~\cite{Borde:1996df,AyonBeato:1998ub,AyonBeato:1999rg,Easson:2002tg,Bronnikov:2005gm,Berej:2006cc,Bronnikov:2012ch,Rinaldi:2012vy,Culetu:2013fsa,Culetu:2014lca,Ghosh:2014pba,Stuchlik:2014qja,Schee:2015nua,Johannsen:2015pca,Dymnikova:2015yma,Myrzakulov:2015kda,Fan:2016hvf,Sebastiani:2016ras,Toshmatov:2017zpr,Chinaglia:2017uqd,Frolov:2017dwy,Pacheco:2018mvs,Simpson:2019mud,Bertipagani:2020awe,Arbey:2021mbl,Nashed:2021pah,Simpson:2021dyo,Franzin:2022iai,Chataignier:2022yic,Ghosh:2022gka,Lewandowski:2022zce,Khodadi:2022dyi,deFreitasPacheco:2023hpb,Farrah:2023opk,Boshkayev:2023rhr,Luongo:2023jyz,Luongo:2023aib,Cadoni:2023lum,Giambo:2023zmy,Bonanno:2023rzk,Cadoni:2023lqe,Luongo:2023xaw,Sajadi:2023ybm,Javed:2024wbc,Ditta:2024jrv,Al-Badawi:2024lvc,Ovgun:2024zmt,Corona:2024gth,Bueno:2024dgm,Konoplya:2024hfg,Pedrotti:2024znu,Bronnikov:2024izh,Kurmanov:2024hpn,Ovalle:2024wtv,Bolokhov:2024sdy,Agrawal:2024wwt,Banerjee:2024sao,Belfiglio:2024wel,Zhang:2024khj,Faraoni:2024ghi,Khodadi:2024efq,Calza:2024qxn,Estrada:2024uuu,KumarWalia:2024yxn,Li:2024ctu,Davies:2024ysj,Estrada:2024moz,Benavides-Gallego:2024hck,Frolov:2024hhe,Balart:2024rtj,Zhang:2024ney,Bueno:2024zsx,Bueno:2024eig,Vertogradov:2025snh,Sajadi:2025prp,Xiong:2025hjn,Estrada:2025aeg,Casadio:2025pun,Dialektopoulos:2025mfz,Harada:2025cwd,Bueno:2025dqk,Fauzi:2025ldu,Kala:2025xnb,Capozziello:2025ycu,Urmanov:2025nou,Alonso-Bardaji:2025qft,Pinto:2025loq,Calza:2025mrt,Bueno:2025gjg,Trivedi:2025vry,Neves:2025uoi,Carr:2025auw,Eichhorn:2025pgy,Ovalle:2025pue,Jusufi:2025qgd,Bueno:2025zaj,Khodadi:2025icd,Trivedi:2025agk,Zare:2025aek,Loc:2025mzc} for a non-exhaustive list of examples. Most of these space-times have been proposed on a phenomenological basis, although a widely held belief is that as of yet unknown quantum gravity effects should ultimately cure the singularities of classical GR. An important consideration for our purposes is that the Ricci scalar of virtually all these regular BHs is, in general, non-vanishing. The previous considerations then lead us to expect that non-minimal couplings of scalar fields to curvature should play an important role in determining the response of regular BHs to scalar perturbations. Yet, to the best of our knowledge, this aspect has been almost completely overlooked in the literature on spectral characteristics of regular BHs.~\footnote{See Refs.~\cite{Crispino:2013pya,Kanti:2014dxa,Pappas:2016ovo,Panotopoulos:2016wuu,Pappas:2017kam,Rincon:2018ktz,MahdavianYekta:2018lnp,Ponglertsakul:2020yrd,Ali:2021gix,Al-Badawi:2023emj} for potential exceptions, not necessarily concerning the specific scalar-curvature coupling we are considering.}

In this pilot study, we take a first step towards closing this gap by investigating how non-minimal couplings affect the spectral properties of regular BHs. We focus on scalar fields, as they provide the simplest testbed: their non-minimal coupling prescription is well-defined and motivated (as argued earlier), their dynamics are easier to handle compared to higher-spin fields, and they capture the essential features of BH responses to external perturbations. Specifically, we focus on classical aspects of the KG equation for massless, non-minimally coupled scalar perturbations around four regular BH space-times with non-vanishing Ricci scalar (the Bardeen, Hayward, Simpson-Visser, and D'Ambrosio-Rovelli BHs), focusing on two cases: a conformally coupled scalar ($\xi=1/6$), and a large negative non-minimal coupling inspired by Higgs inflation ($\xi=-10^4$). We solve the modified Teukolsky equation resulting from the non-minimally coupled KG equation, with the goal of obtaining the corresponding GBFs and Hawking emission spectra.~\footnote{For clarity, throughout the work we use the terms ``Hawking emission'', ``Hawking radiation emission'', and synonyms thereof, to refer specifically to the flux of a hypothetical scalar particle whose action is given by Eq.~(\ref{eq:action}). As we clarify later, we work in the widely adopted test-field regime, where the scalar field is treated as a test perturbation, which does not backreact on the background geometry.} We show that the latter can deviate significantly from their minimally coupled counterparts: these deviations can have a drastic impact on the evaporation history of light primordial regular BHs, and can therefore be particularly important in the scenario where these objects make up at least part of the dark matter in the Universe, by altering their abundance constraints. Interestingly, sizeable differences can arise between the emission spectra of the different BHs themselves (with non-minimal coupling enhancing the emission in some cases and suppressing it in others), pointing towards imprints of the underlying geometry that may allow different BH space-times to be discriminated through their scalar emission. While these results should be viewed as a proof-of-principle, given our focus on scalar emission rather than photon spectra, they constitute an essential first step towards understanding the role of non-minimal couplings in a simple and controlled settings, while paving the way towards future extensions to physically realistic fields.

The rest of this work is then structured as follows. In Sec.~\ref{sec:regular} we present the four regular BHs chosen as case studies. In Sec.~\ref{sec:methodology} we discuss the methodology we adopt to compute the GBFs and scalar Hawking emission spectra of these regular BHs. The resulting spectra are shown and critically discussed in Sec.~\ref{sec:results}. Finally, in Sec.~\ref{sec:conclusions} we provide closing remarks. In what follows, we use the mostly plus signature for the regular BH metrics, and adopt units where $G=c=\hbar=1$.

\section{Regular black holes}
\label{sec:regular}

The Penrose-Hawking singularity theorems establish that, within classical GR, continuous gravitational collapse of matter satisfying the strong or null energy conditions inevitably leads to the formation of space-time singularities, where predictability breaks down~\cite{Penrose:1964wq,Hawking:1970zqf}.~\footnote{See however Refs.~\cite{Sachs:2021mcu,Ashtekar:2021dab,Ashtekar:2022oyq} for an alternative physical interpretation of singularities.} For this reason, singularities are often interpreted as an indication of our incomplete understanding of the physics governing the high-curvature regime. A commonly held view is that quantum gravity effects should ultimately cure these singularities, potentially leading to observable consequences (see e.g.\ Refs.~\cite{Dymnikova:1992ux,Dymnikova:2004qg,Ashtekar:2005cj,Bebronne:2009mz,Modesto:2010uh,Spallucci:2011rn,Perez:2014xca,Colleaux:2017ibe,Nicolini:2019irw,Bosma:2019aiu,Jusufi:2022cfw,Olmo:2022cui,Jusufi:2022rbt,Ashtekar:2023cod,Nicolini:2023hub} for examples of first-principles studies in this direction).

Lacking an universally accepted quantum gravity framework, a different approach towards exploring possible solutions to the singularity problem is more phenomenological in nature. Assuming the validity of a metric description, it is possible to phenomenologically construct space-time geometries which are free from singularities throughout their domain, typically referred to as \textit{regular} BHs (RBHs), see e.g.\ Refs.~\cite{Sebastiani:2022wbz,Lan:2023cvz,Carballo-Rubio:2025fnc} for reviews. Several RBHs and their theoretical and observational properties have been studied over the past decades (see e.g.\ Refs.~\cite{Borde:1996df,AyonBeato:1998ub,AyonBeato:1999rg,Easson:2002tg,Bronnikov:2005gm,Berej:2006cc,Bronnikov:2012ch,Rinaldi:2012vy,Culetu:2013fsa,Culetu:2014lca,Ghosh:2014pba,Stuchlik:2014qja,Schee:2015nua,Johannsen:2015pca,Dymnikova:2015yma,Myrzakulov:2015kda,Fan:2016hvf,Sebastiani:2016ras,Toshmatov:2017zpr,Chinaglia:2017uqd,Frolov:2017dwy,Pacheco:2018mvs,Simpson:2019mud,Bertipagani:2020awe,Arbey:2021mbl,Nashed:2021pah,Simpson:2021dyo,Franzin:2022iai,Chataignier:2022yic,Ghosh:2022gka,Lewandowski:2022zce,Khodadi:2022dyi,deFreitasPacheco:2023hpb,Farrah:2023opk,Boshkayev:2023rhr,Luongo:2023jyz,Luongo:2023aib,Cadoni:2023lum,Giambo:2023zmy,Bonanno:2023rzk,Cadoni:2023lqe,Luongo:2023xaw,Sajadi:2023ybm,Javed:2024wbc,Ditta:2024jrv,Al-Badawi:2024lvc,Ovgun:2024zmt,Corona:2024gth,Bueno:2024dgm,Konoplya:2024hfg,Pedrotti:2024znu,Bronnikov:2024izh,Kurmanov:2024hpn,Capozziello:2024ucm,Ovalle:2024wtv,Bolokhov:2024sdy,Agrawal:2024wwt,Banerjee:2024sao,Belfiglio:2024wel,Zhang:2024khj,Faraoni:2024ghi,Khodadi:2024efq,Calza:2024qxn,Estrada:2024uuu,KumarWalia:2024yxn,Li:2024ctu,Davies:2024ysj,Estrada:2024moz,Benavides-Gallego:2024hck,Frolov:2024hhe,Balart:2024rtj,Zhang:2024ney,Bueno:2024zsx,Bueno:2024eig,Vertogradov:2025snh,Sajadi:2025prp,Xiong:2025hjn,Estrada:2025aeg,Casadio:2025pun,Dialektopoulos:2025mfz,Harada:2025cwd,Bueno:2025dqk,Fauzi:2025ldu,Kala:2025xnb,Capozziello:2025ycu,Urmanov:2025nou,Alonso-Bardaji:2025qft,Pinto:2025loq,Calza:2025mrt,Bueno:2025gjg,Trivedi:2025vry,Neves:2025uoi,Capozziello:2025wwl,Carr:2025auw,Eichhorn:2025pgy,Ovalle:2025pue,Jusufi:2025qgd,Bueno:2025zaj,Khodadi:2025icd,Trivedi:2025agk,Zare:2025aek,Loc:2025mzc} for a non-exhaustive list of examples). While most of these have been introduced on purely phenomenological grounds, we note that several RBHs can be linked to specific non-linear electrodynamics theories~\cite{Bronnikov:2017sgg,Ghosh:2021clx,Bronnikov:2021uta,Bronnikov:2022ofk,Bokulic:2023afx}, although at the price of Laplacian instability around the center~\cite{DeFelice:2024seu}.

In what follows, we briefly describe the four RBH space-times considered in this work, all of which are characterized by a non-vanishing Ricci scalar $R \neq 0$, making the non-minimal coupling of scalar perturbations to curvature discussed earlier particularly relevant. In all four metrics, the parameter $M$ which appears can be unambiguously identified with the (Komar, Arnowitt-Deser-Misner, Misner-Sharp-Hernandez, or Brown-York) mass. As a caveat, we stress that the RBHs we study are regular in that the simplest (second-derivative) curvature invariants $R \equiv g^{\mu\nu}R_{\mu\nu}$, $R_{\mu\nu}R^{\mu\nu}$, and ${\cal K} \equiv R_{\mu\nu\rho\sigma}R^{\mu\nu\rho\sigma}$ are finite. However, this does not necessarily guarantee geodesic completeness of the space-times, with an explicit counterexample being that of the Hayward RBH~\cite{Zhou:2022yio}.~\footnote{Geodesic incompleteness is tied to the non-finiteness of higher-order curvature invariants such as $\Box^nR$ and $R_{\mu\nu\rho\sigma}\Box^nR^{\mu\nu\rho\sigma}$~\cite{Burzilla:2020utr,Giacchini:2021pmr,Antonelli:2025zxh}.} In addition, we note that the stability of RBHs featuring inner horizons is a matter of current debate~\cite{Carballo-Rubio:2021bpr,Carballo-Rubio:2022kad,Bonanno:2022jjp,Bonanno:2022rvo,Carballo-Rubio:2022twq}. Among the RBHs we will study in what follows (which include the Hayward RBH), not all are guaranteed to be geodesically complete. Nevertheless, we ignore the issue of geodesic completeness in our analysis, as our study is phenomenological in spirit: our goal is to assess, within simple, well-behaved models with $R \neq 0$, the impact of non-minimal coupling of scalar perturbations to curvature. As toy models, these metrics remain of considerable interest in the literature, as they provide simple and analytically tractable examples of space-times avoiding curvature singularities, and it is in this spirit that we shall include them in our study, despite these known limitations which the reader should keep in mind.

\subsection{Bardeen black hole}
\label{subsec:bardeen}

The Bardeen metric is easily one of the best known examples of RBHs, and to the best of our knowledge the first one to ever be proposed. It is characterized by the following line element~\cite{Bardeen:1968ghw}:
\begin{eqnarray}
ds^2 &=& -\left ( 1-\frac{2Mr^2}{(r^2 + \ell^2)^{3/2}} \right ) dt^2 + \frac{dr^2}{1-\dfrac{2Mr^2}{(r^2+\ell^2)^{3/2}}} \nonumber \\
&& {}+ r^2d\Omega^2 \,,
\label{eq:ds2bardeen}
\end{eqnarray}
where $d\Omega^2=d\theta^2 +\sin^2(\theta) d\phi^2$ is the metric on the 2-sphere, and as stressed earlier, $M$ is the (Komar, Arnowitt-Deser-Misner, Misner-Sharp-Hernandez, or Brown-York) mass of the BH. Finally, $\ell$ is a regularizing parameter. In particular, for $\ell \neq 0$ the space-time is regular, and features an internal de Sitter (dS) core which replaces the central singularity of the classical Schwarzschild solution. On the other hand, in the $\ell \to 0$ limit the Bardeen space-time recovers the Schwarzschild one. For Eq.~(\ref{eq:ds2bardeen}) to describe a BH, rather than a horizonless compact object, the regularizing parameter $\ell$ is required to satisfy $\ell \leq \ell_{\max}=\sqrt{16/27}M$. While originally introduced on purely phenomenological grounds, it is now known that the Bardeen space-time can originate from a magnetic monopole source~\cite{Ayon-Beato:2000mjt}, potentially within a suitable non-linear electrodynamics theory~\cite{Ayon-Beato:2004ywd}. An alternative possibility for the origin of the Bardeen space-time resides in quantum corrections from modifications to the uncertainty principle~\cite{Maluf:2018ksj}.

\subsection{Hayward black hole}
\label{subsec:hayward}

Alongside the Bardeen BH, the Hayward BH is also among the most well-known examples of RBHs, and is described by the following line element~\cite{Hayward:2005gi}:
\begin{eqnarray}
ds^2 &=& -\left ( 1 - \frac{2Mr^2}{r^3+2M\ell^2} \right ) dt^2 + \frac{dr^2}{1-\dfrac{2Mr^2}{r^3+2M\ell^2}} \nonumber\\
&& {}+r^2d\Omega^2 \,,
\label{eq:ds2hayward}
\end{eqnarray}
where $\ell$ is once more a regularizing parameter, such that for $\ell \neq 0$ the space-time is free of curvature singularities, whereas in the limit $\ell \to 0$ the Schwarzschild BH is recovered. As in the Bardeen case, $\ell \leq \ell_{max}=\sqrt{16/27}M$ is required for the Hayward space-time to describe a BH with an event horizon, rather than a horizonless object. Another common feature between the two metrics is the presence of a dS core, which in the case of the Hayward BH is associated to an effective cosmological constant $\Lambda=3/\ell^2$, preventing the would-be central singularity otherwise present in the Schwarzschild space-time. Like the Bardeen BH, the Hayward BH has been introduced on phenomenological grounds. Nevertheless, various theoretical frameworks have been proposed to explain the origin of this space-time, ranging from modifications to the equation of state of matter at high densities~\cite{Sakharov:1966aja,1966JETP...22..378G}, theoretical finite density and/or curvature proposals related to quantum gravity~\cite{1982JETPL..36..265M,1987JETPL..46..431M,Mukhanov:1991zn}, and corrections arising from theories of non-linear electrodynamics~\cite{Kumar:2020bqf,Kruglov:2021yya}. Despite the generic similarities between the Hayward and Bardeen RBHs, their scalar emission spectra turn out to be rather different, justifying our choice of studying both space-times.

\subsection{Simpson-Visser space-time}
\label{subsec:simpsonvisser}

This metric was introduced by Simpson and Visser as ``the minimal violence to the standard Schwarzschild solution'' needed to enforce regularity, and is one of the best known black-bounce metrics. The line element of the Simpson-Visser space-time is the following:
\begin{eqnarray}
ds^2 &=& - \left ( 1-\frac{2M}{\sqrt{r^2+\ell^2}} \right ) dt^2 + \frac{dr^2}{1-\dfrac{2M}{\sqrt{r^2+\ell^2}}} \nonumber \\
&& {}+ (r^2+\ell^2)d\Omega^2 \,,
\label{eq:ds2simpsonvisser}
\end{eqnarray}
and interpolates between BHs and traversable wormholes (WHs), depending on the value of the regularizing parameter $\ell$, while reducing to the Schwarzschild space-time in the $\ell \to 0$ limit. For $0<\ell/M<2$ the Simpson-Visser metric describes a regular BH with a one-way space-like throat, whereas for $\ell/M=2$ the space-time becomes a one-way WH with an extremal null throat, and for $\ell/M>2$ a traversable WH with a two-way time-like throat. As with several RBH solutions, the Simpson-Visser metric was introduced at a phenomenological level, but can originate from an appropriate theory of non-linear electrodynamics (in this case, one featuring a minimally coupled phantom scalar field)~\cite{Bronnikov:2021uta}. Moreover, it presents characteristic features of several solutions related to quantum gravity, such as its interpolating between BHs and WHs. In our case, in order to deal only with the BH region of parameter space, we set an upper limit on the regularizing parameter $\ell<\ell_{\max}=2M$.

\subsection{D'Ambrosio-Rovelli space-time}
\label{subsec:dambrosiorovelli}

The D'Ambrosio-Rovelli space-time is deeply rooted into Loop Quantum Gravity. Within this framework, it constitutes a natural extension of the Schwarzschild solution, smoothly crossing from the $r=0$ singularity into the interior of a white hole, with the associated black-to-white hole tunneling process being a possible avenue for the resolution of the BH information paradox. Moreover, this space-time can be interpreted as the $\hbar \to 0$ limit of an effective quantum gravity geometry. The D'Ambrosio-Rovelli line element, which resembles the Simpson-Visser one, is given by the following expression~\cite{Bianchi:2018mml,DAmbrosio:2018wgv}:
\begin{eqnarray}
ds^2 &=& - \left ( 1-\frac{2M}{\sqrt{r^2+\ell^2}} \right) dt^2 \nonumber \\
&& {}+ \frac{dr^2}{1-\dfrac{2M}{\sqrt{r^2+\ell^2}}} \left ( 1+\frac{\ell}{\sqrt{r^2 + \ell^2}} \right ) + (r^2+\ell^2)d\Omega^2\,. \nonumber
\label{eq:ds2dambrosiorovelli}
\end{eqnarray}
Following the same arguments offered for the Simpson-Visser space-time, we set an upper limit on the regularizing parameter $\ell<\ell_{\max}=2M$.

\section{Methodology}
\label{sec:methodology}

Here we briefly discuss the necessary steps leading to the computation of the GBFs of the RBHs under consideration, in the presence of a non-minimal coupling of scalar perturbations to curvature. We begin by rewriting the Klein-Gordon equation on a generic spherically symmetric background, in the presence of a non-minimal coupling. This will allow us to work out the appropriate boundary conditions for the radial part of the scalar perturbation, itself a necessary step for proceeding to the GBF computation. We will then proceed to discuss the computation of the GBFs themselves, as well as the resulting emission spectra.

We stress that, in what follows, we work within the test-field regime. That is, the scalar is treated as a test field which does not backreact on the metric, so the RBH space-times are taken as exact solutions of whatever underlying theory they emerge from, an aspect which we remain agnostic about. Within this setup, the non-minimal coupling only alters the KG wave equation, but leaves the underlying space-time unchanged. This justifies the usual temperature definition and flux normalization we adopt in our work. This approach is the standard one in the vast majority of studies on BH spectral characteristics (including GBFs), although we note that a few works going beyond the test-field limit and allowing for backreaction have recently appeared, albeit (so far) within model-dependent frameworks~\cite{Arbey:2025ses,Antoniou:2025bvg}.

\subsection{Non-minimally coupled Klein-Gordon equation on a spherically symmetric background}
\label{subsec:nonminimallykleingordon}

\begin{figure*}[!htb]
\centering
\includegraphics[width=0.49\textwidth]{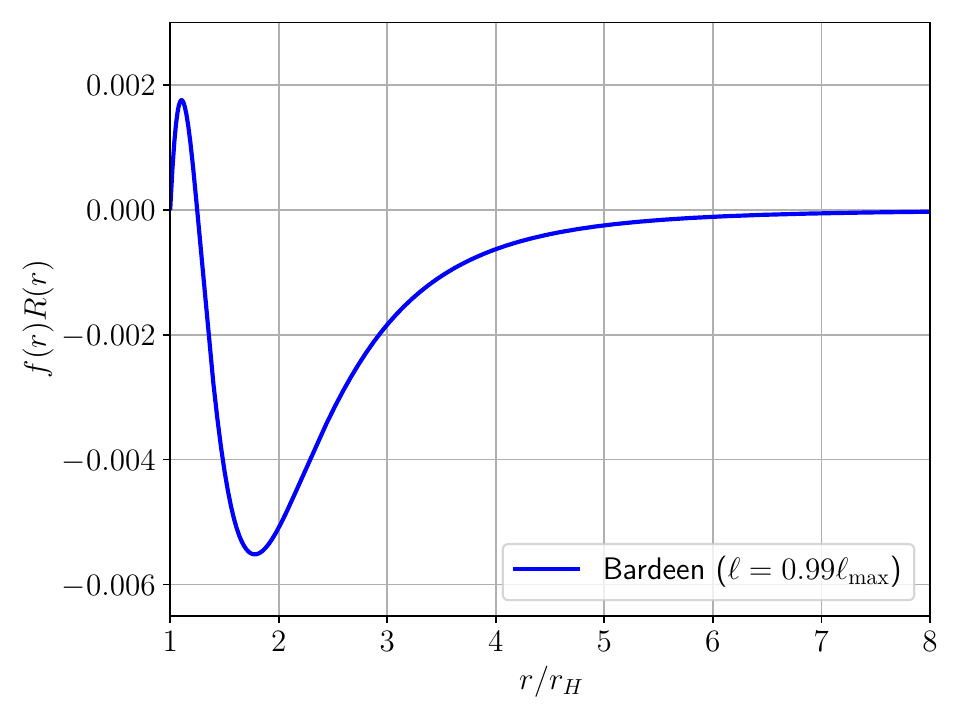}\hfill\includegraphics[width=0.49\textwidth]{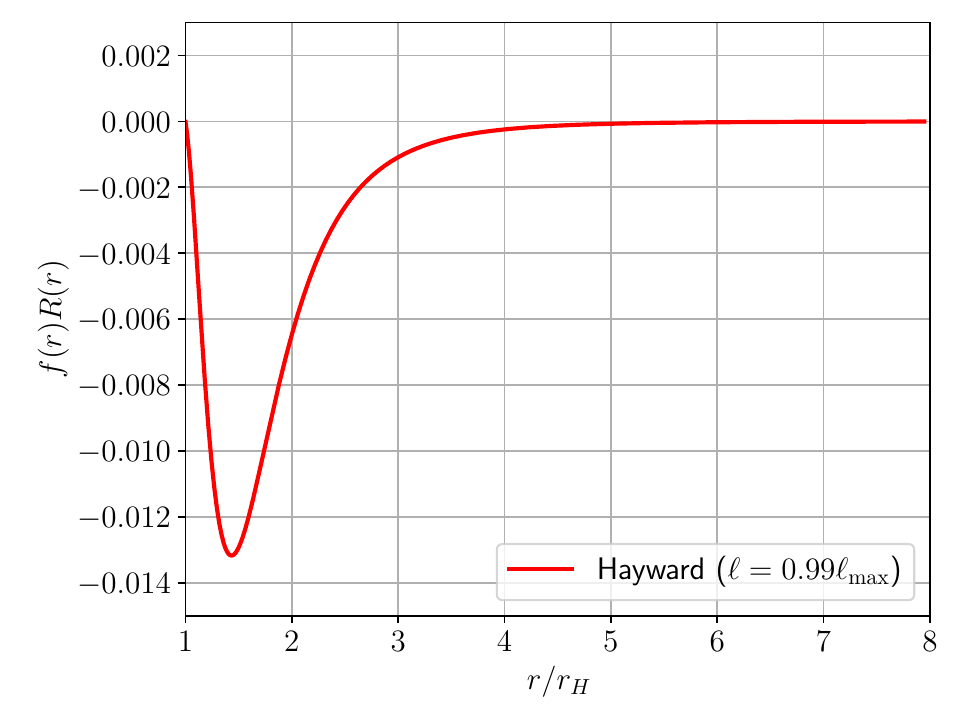} \vskip 0.5cm
\includegraphics[width=0.49\textwidth]{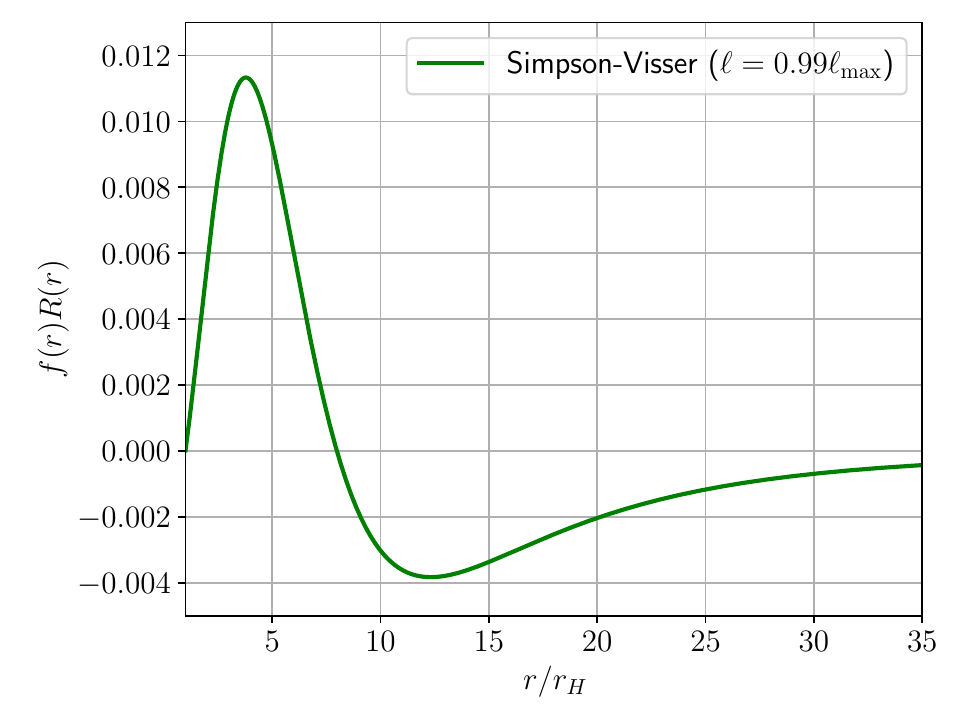}\hfill\includegraphics[width=0.49\textwidth]{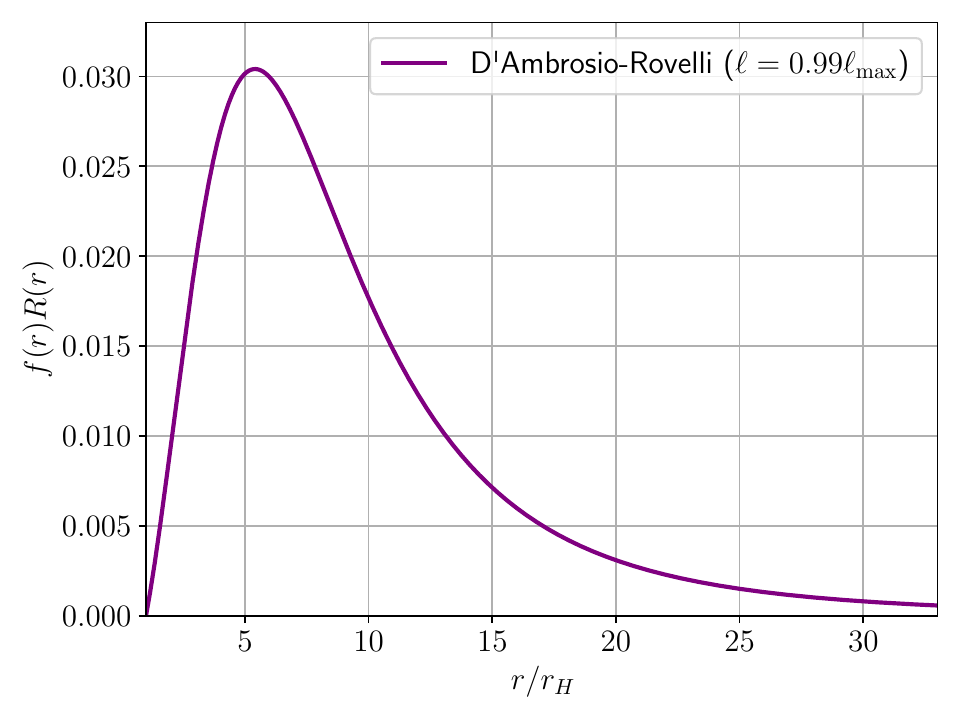}
\caption{Each sub-panel shows the $f(r)R(r)$ factor (with $f(r)$ being minus the $g_{tt}$ metric element in Schwarzschild-like coordinates and $R$ the Ricci scalar) as a function of the radial coordinate normalized by the event horizon radius $r_H$, for the four regular black holes we consider: Bardeen (upper left, blue curve), Hayward (upper right, red curve), Simpson-Visser (lower left, green curve), and D'Ambrosio-Rovelli (lower right, purple curve) regular black holes. In all four cases, we consider a near-extremal value for the regularizing parameter, $\ell=0.99\ell_{\max}$. The $fR$ term is important because, once multiplied by the non-minimal coupling strength $\xi$, it is the one that modifies the standard geometric potential. For all regular black holes considered, $fR \to 0$ both at the event horizon ($r/r_H=1$) and at spatial infinity ($r/r_H \to +\infty$): this implies that the same set of boundary conditions which would be used when computing the graybody factors in the minimally coupled case can be used in the presence of the non-minimal coupling.}
\label{fig:fr}
\end{figure*}

To set the stage, we consider the following class of line elements, manifestly belonging to the Petrov type-D class of metrics:
\begin{equation}
ds^2=-f(r)dt^2+g(r)^{-1}dr^2+h(r)d\Omega^2\,.
\label{eq:ds2}
\end{equation}
It is evident that all four RBHs we are considering have line elements which are described by Eq.~(\ref{eq:ds2}), for appropriate choices of $f(r)$, $g(r)$, and $h(r)$. We require asymptotic flatness and therefore impose the following constraints on these functions:
\begin{equation}
f(r) \xrightarrow{r \to \infty} 1\,, \quad g(r) \xrightarrow{r \to \infty} 1\,, \quad h(r) \xrightarrow{r \to \infty} r^2\,.
\label{eq:asflat}
\end{equation}
These conditions are met by the four RBHs we are considering in the present work.

On this background, it is easy to show that the non-minimally coupled KG equation, given in Eq.~(\ref{eq:kgm0}), takes the following form:
\begin{equation}
\begin{split}
&\frac{\csc^2(\theta)\partial^2_\varphi\phi+\cot(\theta)\partial_\theta \phi+\partial^2_\theta\phi+gh'\partial_r\phi}{h}+\\
&\frac{(gf'+g'f)\partial_r\phi+2fg\partial^2_r\phi-2\partial^2_t\phi}{2f}-\xi R\phi=0\,,
\end{split}
\label{eq:nmkg}
\end{equation}
where $'$ denotes a derivative with respect to $r$ (we note that the functions $f$, $g$, and $h$ depend only on $r$), and the Ricci scalar $R(r)$ is given by the following:
\begin{equation}
\begin{split}
R(r)=&\frac{1}{2f^2h^2}\Big(gh^2f'^2 \\
&-fh[2gf'h'+h(f'g'+2gf'')] \\
&-f^2[2h(-2+g'h'+2gh'')-gh'^2]\Big)\,.
\end{split}
\end{equation}
Direct calculation shows that Eq.~(\ref{eq:nmkg}) is separable if we make the following ansatz for the scalar perturbation:
\begin{equation}
\phi(t,r,\theta,\phi)=T(t)\psi(r)S(\theta,\varphi)\,.
\end{equation}
In this case, the time dependence of the perturbation is harmonic, corresponding to a monochromatic mode $T(t)=e^{i\omega t}$. On the other hand, the angular part satisfies the standard spherical harmonics equation, given by the following expression:
\begin{equation}
\left [ \partial^2_\theta+\csc(\theta)^2\partial^2_\varphi+\cot(\theta)\partial_\theta \right ] S=-l(l+1)S\,,
\end{equation}
where $l$ is the multipole number, not to be confused with $\ell$, the regularizing parameter. With regards to the radial part of the perturbation, it is convenient to work with the tortoise coordinate $r_{\star}$, which, for metrics of the form of Eq.~(\ref{eq:ds2}), is defined as follows:
\begin{equation}
\frac{dr_{\star}}{dr}=\frac{1}{\sqrt{f(r)g(r)}}\,.
\label{eq:rstar}
\end{equation}
Expressed in terms of the tortoise coordinate, the equation for the radial part of the perturbation takes the following form:
\begin{equation}
(\partial_{r_{\star}}^2+\omega^2-V_{\text{eff}})\psi(r_{\star})=0\,,
\label{eq:schrodingerkleingordon}
\end{equation}
where the derivative with respect to the tortoise coordinate is related to the standard radial derivative by $\partial_{r_{\star}}\equiv\sqrt{fg}\partial_r$. $V_{\text{eff}}$ appearing in Eq.~(\ref{eq:schrodingerkleingordon}) is an effective potential given by the following:
\begin{equation}
V_{\text{eff}}=V_l+\xi fR\,,
\label{eq:effective}
\end{equation}
with $V_l$ being the usual geometric potential for a minimally coupled scalar field, which reads as follows:
\begin{equation}
V_l=l(l+1)\frac{f}{h}-\frac{1}{4h} \left [ \frac{fg}{h}(h'^2-2hh'')-h'(f'g+fg') \right ] \,.
\label{eq:vlgeometricpotential}
\end{equation}
We have suppressed an obvious $r_{\star}$ and/or $r$ dependence in Eqs.~(\ref{eq:effective},\ref{eq:vlgeometricpotential}). In Fig.~\ref{fig:fr} we plot the term $f(r)R(r)$ for the four RBHs we consider, as a function of the radial coordinate, normalized by the event horizon radius $r_H$. This term is important because it is precisely the one that modifies the standard geometric potential when the non-minimal coupling is switched on, as shown in Eq.~(\ref{eq:effective}): it will therefore play a key role in shaping the effective potential barrier and, consequently, the resulting GBFs. Depending on the distance from the event horizon and on the sign of $\xi$, this term can either enhance or suppress the geometric potential. For instance, in the Bardeen (upper left panel) and Simpson-Visser (lower left panel) cases, it is positive near the horizon, changes sign at intermediate radii, and asymptotically approaches zero. In contrast, for the Hayward BH (upper right panel) the term remains negative throughout, while for the D'Ambrosio-Rovelli BH (lower right panel) it remains positive. In both cases, this term asymptotically vanishes.

A key observation is that the term  $\xi fR$ enjoys the same well-behaved characteristics of $V_l$. Specifically, it vanishes both at the event horizon $r_H$ ($r_{\star} \to -\infty$), since $f(r_H)=0$ and $R(r \to r_H)$ is finite, and at spatial infinity ($r_{\star} \to +\infty$), since asymptotic flatness requires $f(r \to +\infty) \to 1$ and $R(r \to +\infty) \to 0$. For all four RBHs, these features are evident from Fig.~\ref{fig:fr}, and imply that we can express the asymptotic solutions to Eq.~(\ref{eq:schrodingerkleingordon}) in the following form:
\begin{align}
\psi(r_{\star}\rightarrow -\infty)
&= \mathfrak{a}e^{i\omega r_{\star}}+\mathfrak{b}e^{-i\omega r_{\star}}\,,
\label{eq:asymptminus}\\
\psi(r_{\star}\rightarrow +\infty)
&= ae^{i\omega r_{\star}}+be^{-i\omega r_{\star}}\,,
\label{eq:asymptplus}
\end{align}
where $\mathfrak{a}$, $\mathfrak{b}$, $a$, and $b$ are free parameters. The fact that the asymptotic solutions for the radial part of the perturbation are those of Eqs.~(\ref{eq:asymptminus},\ref{eq:asymptplus}) is important because it ensures that, despite the presence of a non-minimal coupling, the perturbation behaves regularly at both boundaries. This implies that, when computing the GBFs, we can set the same standard boundary conditions we would use in the absence of the non-minimal coupling.

\subsection{Graybody factors and Hawking radiation emission spectra}
\label{subsec:gbf}

\begin{figure*}[!htb]
\centering
\includegraphics[width=0.49\textwidth]{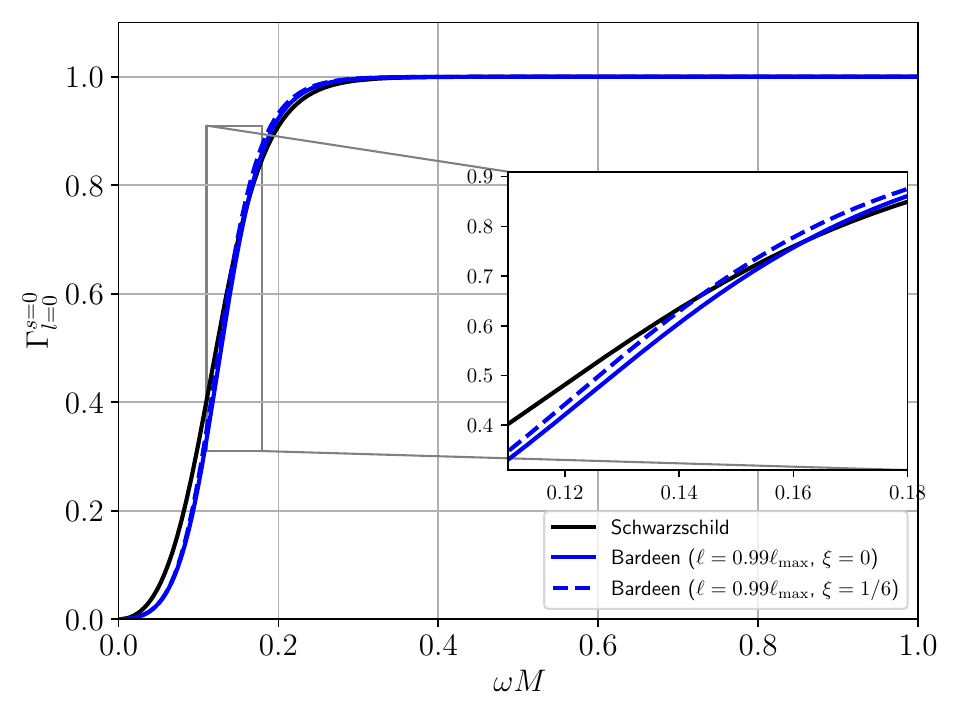}\hfill\includegraphics[width=0.49\textwidth]{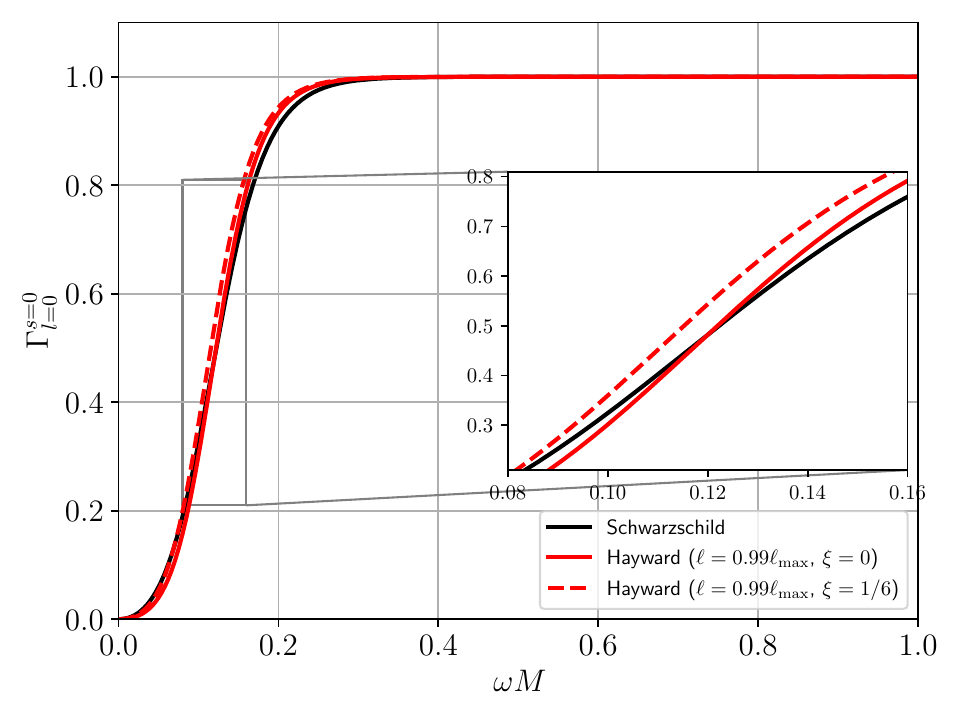} \vskip 0.5cm
\includegraphics[width=0.49\textwidth]{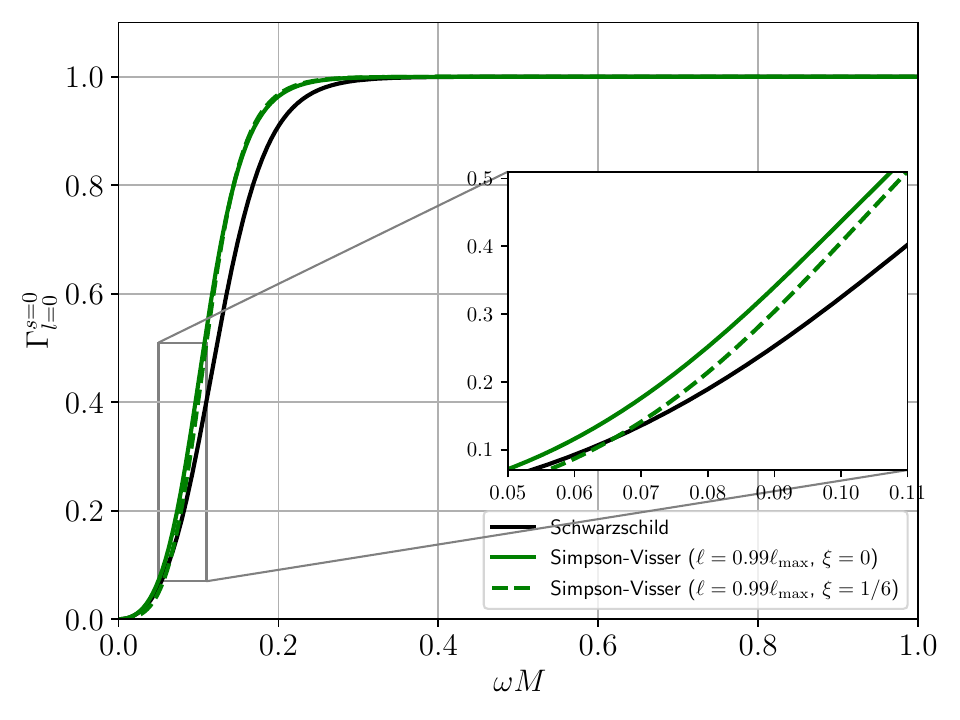}\hfill\includegraphics[width=0.49\textwidth]{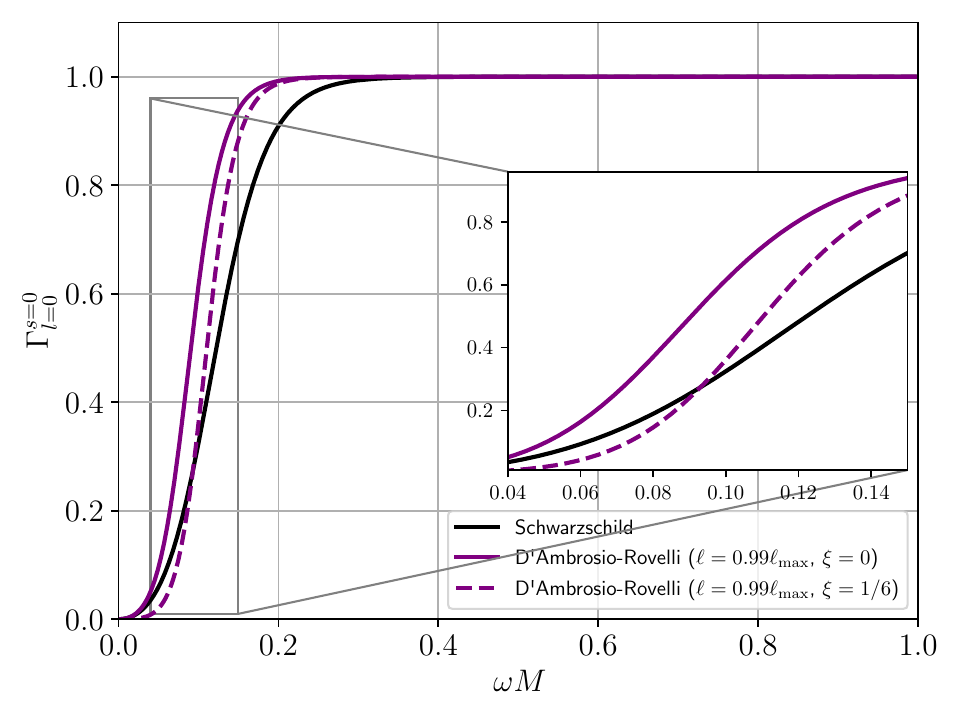}
\caption{Each sub-panel shows the graybody factors $\Gamma_{l=0}^{s=0}$, as a function of $\omega M$, for the four regular black holes we consider: Bardeen (upper left, blue curve), Hayward (upper right, red curve), Simpson-Visser (lower left, green curve), and D'Ambrosio-Rovelli (lower right, purple curve) regular black holes. In each sub-panel, we compare the GBFs for the four minimally coupled regular black holes ($\xi=0$, colored solid curves) to their conformally coupled counterparts ($\xi=1/6$, colored dashed curves), and the Schwarzschild GBF (black curve). The insets zoom into the regions where the deviations between the three GBFs are most significant. For illustrative purposes we only plot $\Gamma_{l=0}^{s=0}$, since we are interested in scalar emission ($s=0$) and the dominant emission mode is the $l=0$ one. We have fixed the regularizing parameter to the near-extremal value $\ell=0.99\ell_{\max}$ in all four cases. The features shown in this plot do not change sensibly for higher values of $l$ and other values of $\ell$.}
\label{fig:gbf_conformal}
\end{figure*}

\begin{figure*}[!htb]
\centering
\includegraphics[width=0.63\textwidth]{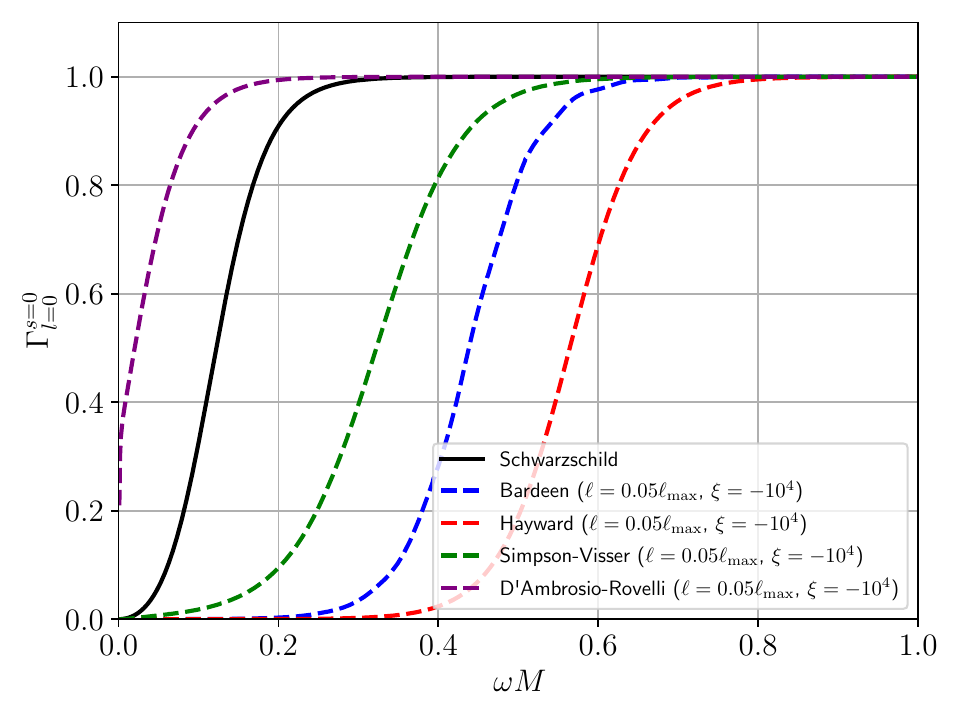}\
\caption{Graybody factors $\Gamma_{l=0}^{s=0}$ for the four regular black holes we consider, in the inflationary-inspired case where $\xi=-10^4$, and setting the regularizing parameter to $\ell=0.05\ell_{\max}$. The blue, red, green, and purple dashed curves correspond to the Bardeen, Hayward, Simpson-Visser, and D'Ambrosio-Rovelli regular black holes respectively, whereas the black solid curve corresponds to the Schwarzschild black hole, which here is taken as the comparison baseline instead of the respective minimally coupled counterparts: the reason is that the dynamical range of the resulting Hawking emission spectra is much larger than in the conformal case, and the corresponding minimally coupled spectra are nearly indistinguishable from each other and from the Schwarzschild spectrum across the plotting range we will later adopt.}
\label{fig:gbf_inflation}
\end{figure*}

We now have all the tools in place to compute the (scalar) GBFs for the four RBHs considered. We recall that GBFs are functions of energy and angular momentum which control the deviation of the emitted Hawking radiation spectrum from that of a blackbody (see e.g.\ Ref.~\cite{Sakalli:2022xrb} for a recent review). Although at the event horizon the Hawking radiation spectrum is exactly of the blackbody form, the radiation propagating outwards and reaching the observer is modified by the curvature-induced potential barrier surrounding the BH. The GBFs are then none other than the frequency-dependent transmission functions $\Gamma_l^s(\omega)$, with $\omega$ the energy/frequency of the particle of spin $s$ being emitted, and $l$ its angular momentum/orbital number.

To determine the GBFs we set up a classical scattering problem with waves originating from the horizon (due to the symmetries of the problem, this is equivalent to having waves incoming from spatial infinity, which would be the more natural choice for scattering problems). Mathematically speaking, we set up purely ingoing boundary conditions, while normalizing the wave function to unity, which corresponds to setting $\mathfrak{a}=0$ and $\mathfrak{b}=1$ in Eq.~(\ref{eq:asymptminus}). With these boundary conditions, the evolution of the perturbations is then governed by Eq.~(\ref{eq:schrodingerkleingordon}), which represents a modified Teukolsky equation~\cite{Teukolsky:1973ha} for non-minimally coupled scalar perturbations. This is solved by integrating Eq.~(\ref{eq:schrodingerkleingordon}) from the near-horizon solution [Eq.~(\ref{eq:asymptminus}) with $\mathfrak{a}=0$ and $\mathfrak{b}=1$] out to the far-horizon region, to determine the transmission and reflection coefficients, which are given by the following:
\begin{equation}
T=\Gamma_l(\omega)=\frac{1}{\vert b \vert^2}\,, \quad R=\frac{1}{\vert a \vert^2}\,,
\end{equation}
with the transmission coefficient $T$ being precisely the GBF we are interested in. In practice, for each of the four RBHs considered, we compute the GBFs numerically, by making use of a suitably modified version of the publicly available \texttt{GrayHawk} code~\cite{Calza:2025whq}, where the term $\xi fR$ has been added to the effective potential as in Eq.~(\ref{eq:effective}). For further details, we refer the reader to Ref.~\cite{Calza:2025whq} and Appendix~A of Ref.~\cite{Calza:2024fzo}, noting that a similar methodology has been applied in several other works recently (see e.g.\ Refs.~\cite{Rosa:2012uz,Rosa:2016bli,Calza:2021czr,Calza:2022ioe,Calza:2022ljw,Calza:2023rjt,Calza:2023gws,Calza:2023iqa,Calza:2024xdh,Arbey:2025dnc,Yuan:2025eyi,Lutfuoglu:2025ljm,Calza:2025mwn,Lutfuoglu:2025bsf,Bonanno:2025dry}). We compute the GBFs for minimally coupled ($\xi=0$) and non-minimally coupled ($\xi \neq 0$) scalars ($s=0$), for angular momentum up to $l=4$, verifying that our results do not change significantly if we include higher $l$ modes.

For what concerns the non-minimal coupling, we are interested in two specific case studies. The first is the theoretically well-motivated conformal coupling $\xi=1/6$, providing a natural benchmark for comparison to the minimally coupled case. In this case, for all four RBHs, we consider a near-extremal regularizing parameter, $\ell=0.99\ell_{\max}$, in order to amplify the differences relative to the minimally coupled case. In passing we note that it does not make sense to take the extremal case $\ell=\ell_{\max}$, since in that case the BH has zero temperature (the reason being that the inner and outer horizons coincide, leading to zero surface gravity), and therefore there is no Hawking radiation. 

The second case is  motivated by Higgs inflation, where $\xi$ is large and negative. In particular, we take $\xi=-10^4$, in agreement with the value required to match the observed spectral index within Higgs inflation~\cite{Bezrukov:2007ep}. In this case, we note that the Hawking radiation spectrum can deviate significantly from the minimally coupled case even for RBHs far from extremality. With this in mind, we therefore set $\ell=0.05\ell_{\max}$. For simplicity, we refer to this case as ``inflation-inspired'', even though we stress that this in principle has nothing to do with inflation, nor are the scalar perturbations in any way related to the inflaton. Summarizing, we consider two case studies:
\begin{enumerate}
\item \textit{\textbf{conformal case}}: $\xi=1/6$, $\ell=0.99\ell_{\max}$;
\item \textit{\textbf{inflationary-inspired case}}: $\xi=-10^4$, $\ell=0.05\ell_{\max}$.
\end{enumerate}
This choice allows us to probe two distinct phenomenological cases, as the opposite signs of the non-minimal coupling give rise to qualitatively distinct scalar-curvature interactions, although we stress that the inflationary-inspired case at the very least raises questions on potential violations of unitarity, as discussed in the literature in the context of Higgs inflation~\cite{Giudice:2010ka,Calmet:2013hia,Fumagalli:2017cdo}.

In Fig.~\ref{fig:gbf_conformal} we show the $\Gamma_{l=0}^{s=0}$ GBFs for the four RBHs that we consider in the conformal case (one for each sub-panel). We focus on the s-wave ($l=0$) mode since it dominates the emitted Hawking radiation. In all four cases, we compare the GBFs for the minimally coupled, near-extremal RBHs ($\ell=0.99\ell_{\max}$, $\xi=0$, colored solid curves) to their conformally coupled counterparts ($\ell=0.99\ell_{\max}$, $\xi=1/6$, colored dashed curves), as well as the Schwarzschild GBF, which is recovered in the $\ell \to 0$ limit. We immediately notice an important qualitative difference between two classes of RBHs. The Bardeen and Hayward cases (upper left and upper right sub-panels) fall in the first class: in this case, the non-minimally coupled GBF is consistently (slightly) enhanced compared to its minimally coupled counterpart. The reason can be appreciated from the respective sub-panels in Fig.~\ref{fig:fr}. Throughout most of the region relevant to the scattering problem, for $\xi>0$ the $\xi fR$ correction to the geometric potential is negative. This lowers the potential barrier, thereby facilitating the transmission of radiation, increasing the transmission factor, and therefore the GBF. The converse occurs for the Simpson-Visser and D'Ambrosio-Rovelli cases (lower left and lower right sub-panels), where the potential barrier is instead raised, as can be appreciated from the respective sub-panels in Fig.~\ref{fig:fr}: this leads to an overall suppression of the GBF. Moreover, unlike the case of Bardeen and Hayward BHs, whose GBFs are very close to the Schwarzschild one regardless of the value of $\xi$, both the Simpson-Visser and D'Ambrosio-Rovelli GBFs plateau at slightly lower energies compared to the Schwarzschild case.

In Fig.~\ref{fig:gbf_inflation} we instead show the $\Gamma_{l=0}^{s=0}$ GBFs for the same four RBHs, in the inflationary-inspired case ($\xi=-10^4$, $\ell=0.05\ell_{\max}$). A posteriori, we will find that the dynamical range of the resulting Hawking emission spectra is much larger than in the conformal case, and the corresponding minimally coupled spectra are nearly indistinguishable from each other and from the Schwarzschild spectrum across the plotted range. Therefore, for the emission spectra of the four RBHs in the inflationary-inspired case, the comparison baseline will be the Schwarzschild spectrum. A similar line of reasoning is applied to the GBFs, and therefore the comparison baseline in Fig.~\ref{fig:gbf_inflation} is the $\Gamma_{l=0}^{s=0}$ Schwarzschild GBF. We observe that for the Bardeen (blue curve), Hayward (red curve), and Simpson-Visser (green curve) cases, the GBFs are overall strongly suppressed relative to their Schwarzschild counterpart: or, stated, differently, they start rising at much larger energies, $\omega M \gtrsim 0.2$, at which point the Schwarzschild GBF has already plateaued at $\approx 1$. The reason is that, with a large negative $\xi$, the regions where the factor $fR$ is negative (see Fig.~\ref{fig:fr}) give rise to a large positive contribution to the effective potential: this forms a localized barrier-like feature which lowers the transmission probability, and thus the GBFs. Among the three, the suppression is largest for the Hayward case, since the $fR$ factor is negative throughout the entire space-time, leading to a large enhancement of the potential barrier. These features are expected to reduce the transmissive window in frequency space for these three RBHs, leading to an overall suppression of their Hawking emission spectra: such an expectation will be confirmed in our analysis.

The behaviour of the D'Ambrosio-Rovelli GBF is instead completely different. In this case, the GBF rises quickly already at very low energies ($\omega/M \lesssim 0.02$), and plateaus to $\approx 1$ at significantly lower energies compared to its Schwarzschild counterpart (although this is far from obvious on a linear plot, given its steep rise, the D'Ambrosio-Rovelli GBF vanishes for $\omega M=0$, as we have explicitly checked). This enhancement is directly tied to the fact that the $fR$ factor is positive throughout the entire space-time. When multiplied by a large negative value of $\xi$, the (negative) contribution of the non-minimal coupling to the effective potential dominates over the geometrical one. As a result, $V_{\text{eff}}$ features a deep, large well, which enhances the transmission probability, especially at low energies. Conversely to the Bardeen, Hayward, and Simpson-Visser RBHs, this feature is expected to increase the transmissive window in frequency space for the D'Ambrosio-Rovelli BH, leading to an overall enhancement of the resulting Hawking evaporation spectrum, expectation which our later analysis will confirm.

Aside from the GBFs, a key quantity which characterizes the emitted Hawking radiation is the BH temperature $T$. In what follows, we adopt the standard Gibbons-Hawking prescription for computing the temperatures of the four RBHs considered. In practice, this involves performing a Wick rotation of the metrics and imposing regularity in the Euclidean period: this definition directly links the temperature to the BH surface gravity $\kappa$, in agreement with Hawking's original derivation~\cite{Hawking:1970zqf,Hawking:1975iha} and the results obtained via Bogoliubov transformations between in- and out-vacua.~\footnote{We note that alternative prescriptions for computing the temperature of RBHs have been discussed in the literature, see e.g.\ Refs.~\cite{Hayward:2008jq,Maluf:2018lyu} (see also Refs.~\cite{Faraoni:2021jri,Gallerani:2024gdy,Faraoni:2025alq}). Nevertheless, the Gibbons-Hawking definition we adopt is consistent with our boundary conditions, and the fact that the non-minimal coupling enters only the wave equation, leaving the horizon geometry (and therefore the surface gravity) unchanged. As a caveat, we note that the identification between cyclic imaginary time and temperature underlying the Gibbons-Hawking approach implicitly assumes the standard Boltzmann-Gibbs distribution, but may be inconsistent in the presence of alternative entropic frameworks, themselves potentially tied to theories beyond GR~\cite{Nojiri:2021czz,Nojiri:2022sfd,Nojiri:2022ljp,Lu:2024ppa,Nojiri:2024zdu}. Here we make the choice of remaining agnostic about the theoretical origin of the RBHs we consider, and therefore set aside these statistical-mechanical issues.} Under these standard assumptions, the temperature of the RBHs we are considering is given by the following:
\begin{eqnarray}
T=\sqrt{\frac{g(r)}{f(r)}}\frac{f'(r)}{4\pi} \Bigg\vert_{r_H}\,,
\label{eq:temperature}
\end{eqnarray}
with the expression evaluated at the event horizon. The temperatures of the four RBHs as a function of the regularizing parameter can be found in earlier plots by some of us, specifically Fig.~1 of Ref.~\cite{Calza:2024fzo} and Fig.~1 of Ref.~\cite{Calza:2024xdh}, and will therefore not be shown here. In all cases, however, we note that the temperature is a monotonically decreasing function of $\ell$, a feature common to nearly all RBHs.

Once the GBFs and temperatures of the RBHs we study are known, we can compute their Hawking radiation emission rate. We stress that here we are specifically concerned with the emission of \textit{scalar particles} with energies $E_s$, \textit{not} photons, as we have considered a scalar-curvature non-minimal coupling rather than a vector-curvature one. Under the reasonable assumption that the emitted scalar particles do not couple directly to the regularizing parameter $\ell$, the number of scalars with energy $E_s$ emitted per unit time per unit energy is then given by the following:
\begin{eqnarray}
\frac{d^2N}{dtdE_s}(E_s,\ell)=\frac{1}{2\pi}\sum_{l}(2l+1)\frac{\Gamma_{l}^{s=0}(E_s,\ell)}{e^{E_s/T(\ell)}-1}\,,
\label{eq:d2ndtdei}
\end{eqnarray}
where the degeneracy factor of $(2l+1)$ corresponding to the $m$-modes for a given $l$ follows from spherical symmetry, and we have implicitly set $k_B=1$.

\section{Results}
\label{sec:results}

\begin{figure*}[!htb]
\centering
\includegraphics[width=0.49\linewidth]{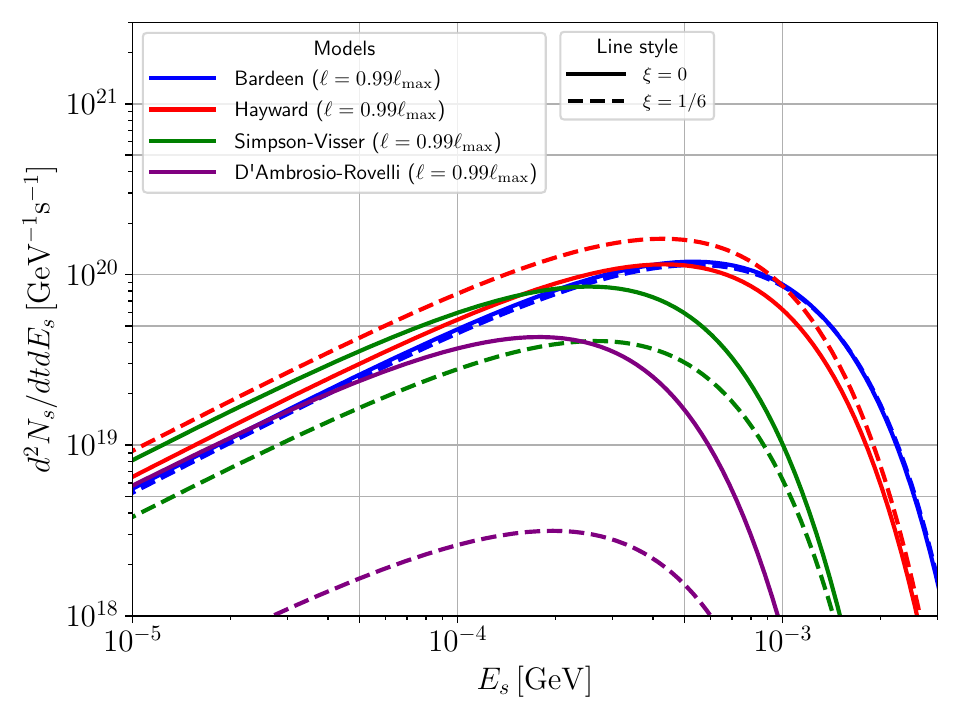}\hfill\includegraphics[width=0.49\linewidth]{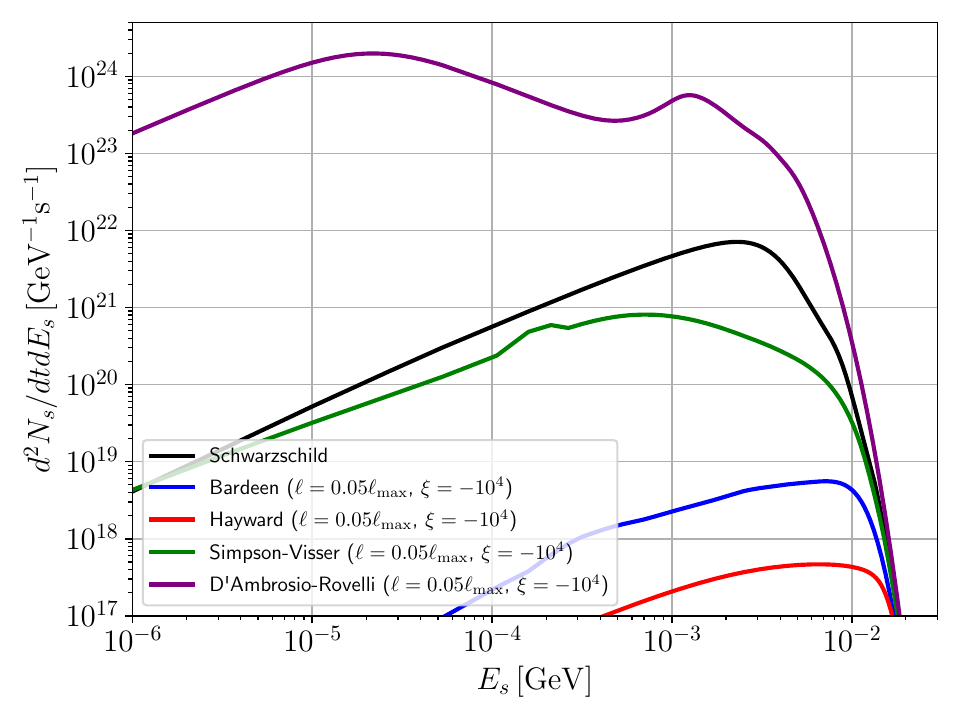}
\caption{Scalar Hawking emission spectra resulting from the evaporation of a regular black hole of mass $M=10^{16}\,{\text{g}}$. \textit{Left panel}: spectra computed in the conformal case, where $\xi=1/6$ and $\ell=0.99\ell_{\max}$, for the Bardeen (blue curves), Hayward (red curves), Simpson-Visser (green curves), and D'Ambrosio-Rovelli (purple curves) regular black holes. In all cases, the resulting spectra in the non-minimally coupled cases (dashed curves) are compared against their minimally coupled counterparts (solid curves). \textit{Right panel}: spectra computed in the inflationary-inspired case, where $\xi=-10^4$ and $\ell=0.05\ell_{\max}$, for the Bardeen (blue curve), Hayward (red curve), Simpson-Visser (green curve), and D'Ambrosio-Rovelli (purple curve) regular black holes. In this case, the comparison baseline is the emission spectrum of the Schwarzschild black hole (black curve): the reason is that in this regime ($\vert \xi \vert \gg 1$) the dynamical range of the resulting spectra is much larger than in the conformal case, and the corresponding minimally coupled spectra are nearly indistinguishable from each other, and from the Schwarzschild spectrum, across the plotted range.}
\label{fig:spectra}
\end{figure*}

We can now compute the Hawking emission spectra for the two case studies we consider (conformal case and inflationary-inspired case). The resulting spectra for a BH of mass $M=10^{16}\,{\text{g}}$ are shown in Fig.~\ref{fig:spectra}, with the conformal (inflationary-inspired) case given in the left (right) panel.~\footnote{For this choice of mass, the evaporation time is much larger than the age of the Universe and the BHs have lost a negligible fraction of their mass from formation up to the present time.} For the conformal case, we compare the four spectra accounting for the non-minimal coupling (dashed curves) to their four minimally coupled counterparts (solid curves). For the inflationary-inspired case, the four spectra in the non-minimally coupled case are instead all compared to the Schwarzschild emission spectrum: the reason is that in this regime ($\vert \xi \vert \gg 1$) the dynamical range of the resulting spectra is much larger than in the conformal case, and the corresponding minimally coupled spectra are nearly indistinguishable from each other (and from the Schwarzschild spectrum, given the choice of $\ell=0.05\ell_{\max}$) across the plotted range.

We start by discussing the conformal case (left panel). In the case of Bardeen RBHs (blue curves) the conformally coupled case is virtually indistinguishable from its minimally coupled counterpart. This agrees with the fact that the corresponding GBFs are nearly indistinguishable (see the upper left sub-panel of Fig.~\ref{fig:gbf_conformal}). For Hayward RBHs (red curves) we instead see an enhancement of the emission spectrum in the non-minimally coupled case, which is larger by about a factor of $2$ compared to its minimally coupled counterpart. This is in good agreement with the features observed in the corresponding GBFs (see the upper right sub-panel of Fig.~\ref{fig:gbf_conformal}), in particular the fact that the non-minimally coupled GBF remains larger than the minimally coupled one \textit{at all energies}, particularly in the low-energy regime which dominates the emission spectrum. This enhancement directly reflects the fact that the non-minimal coupling lowers the potential barrier, since the $\xi fR<0$ factor is negative throughout the entire space-time.

Very different features are instead observed in the Simpson-Visser (green curves) and D'Ambrosio-Rovelli (purple curves) cases. In both cases, the emission spectra in the conformally coupled case are suppressed relative to their minimally coupled counterparts. This agrees with the observed suppression of the GBFs (see the lower left and lower right sub-panels of Fig.~\ref{fig:gbf_conformal}). While the suppression is relatively mild in the Simpson-Visser case (no more than a factor of $2$), it is instead quite significant in the D'Ambrosio-Rovelli case, where around the peak ($E_s \sim 10^{-4}\,{\text{GeV}}$) it exceeds an order of magnitude. Again, these features are consistent with those observed in the GBFs, reflecting the fact that in the D'Ambrosio-Rovelli case the non-minimal coupling significantly raises the potential barrier, since $\xi fR>0$ throughout the entire space-time (whereas in the Simpson-Visser case this is no longer true sufficiently far from the event horizon).

We now turn to the inflationary-inspired case, where we recall that the baseline for comparison is the Schwarzschild spectrum. In this case, we find that the Bardeen (blue curve), Hayward (red curve), and Simpson-Visser (green) spectra are all suppressed relative to the Schwarzschild one, by more than four order of magnitude in the Hayward case. This confirms the expectation stated earlier, related to the fact that the associated GBFs (see Fig.~\ref{fig:gbf_inflation}) start rising and plateau at much higher energies compared to their Schwarzschild counterpart, for reasons tied to the sign of $\xi fR$. The Simpson-Visser and Bardeen spectra also display a bump in the low-energy part of their spectra, for energies $E \gtrsim 10^{-4}\,{\text{GeV}}$ (although this is significantly less evident in the Bardeen case). This feature is a result of the sign change in the $fR$ factor. For the Hayward case, where the $fR$ factor is negative throughout the entire space-time, this sign change does not occur, explaining the absence of bumps in the emission spectrum.

The most interesting case is the D'Ambrosio-Rovelli one. Here we observe that the spectrum is significantly enhanced compared to the Schwarzschild case, by up to five orders of magnitude. This is consistent with the fact that the GBF starts rising at much lower energies compared to its Schwarzschild counterpart (see Fig.~\ref{fig:gbf_inflation}), for reasons discussed earlier and related to the sign of $\xi fR$. To be clear, this significant enhancement in the Hawking radiation in the inflationary-inspired D'Ambrosio-Rovelli case does not signal any violation of unitarity. Instead, what happens is that the transmissive window is broadened in frequency space. Stated differently, the convolution of the GBFs with the blackbody spectrum extends over a much wider energy range, leading to an overall enhancement in the spectrum.

Remaining in the D'Ambrosio-Rovelli case, we also notice a peak at higher energies, $E \sim 10^{-3}\,{\text{GeV}}$. The reason behind this has to do once more with the deep well in $V_{\text{eff}}$ associated to the large negative value of $\xi$ in combination with the positive $fR$ factor. The well is deeper and larger for lower $l$ modes, which in the scattering problem characterize a higher transmission probability for low frequencies. This implies that the contributions of different $l$ modes to the spectrum peak at different energies: specifically, the contributions from lower $l$ modes are larger and peak at lower energies. As a result, in the spectrum we can appreciate not only the peak due to the $l=0$ mode at low energies, $E \gtrsim 10^{-5}\,{\text{GeV}}$, but also those due to higher $l$ modes. These features are not present in the other three space-times, as they are directly tied to the overall positivity of the $fR$ factor.

Before closing, we briefly comment on potential observational implications of our findings. Direct detection of Hawking radiation has yet to be achieved,~\footnote{Hawking radiation has been observed in analogue BH models such as those based on atomic Bose Einstein-condensates. However, by ``direct detection'' we refer to the actual detection of Hawking radiation from astrophysical or primordial BHs, which is well out of reach at the time of writing.} and no scalar component has ever been observed, so any discussion in this direction is necessarily speculative. Nevertheless, the large differences in the emission spectra we observe in the inflationary-inspired scenario (in particular the strong enhancement we find in the D'Ambrosio-Rovelli case) suggest that, if light scalar fields exist and couple (however weakly) to ordinary matter, their emission could significantly affect the total emission spectra from light primordial RBHs. In fact, such an emission would directly affect the evaporation history of these objects, and therefore their abundance constraints. These considerations can therefore be particularly important in the case where light, evaporating primordial RBHs make up at least part of the dark matter in the Universe (see e.g.\ Refs.~\cite{Calza:2024fzo,Calza:2024xdh,Calza:2025mwn}). Moreover, the qualitative differences in the emission spectra (e.g.\ the bumps in the D'Ambrosio-Rovelli and Bardeen cases due to the sign inversion in $fR$) could in principle provide a way to discriminate between different RBH geometries. In the conformally coupled scenario, these differences are of order unity and therefore much more challenging to detect, except again in the D'Ambrosio-Rovelli case, where the conformal coupling suppresses the emission spectrum by an order of magnitude. At any rate, it goes without saying that detecting the effects shown in Fig.~\ref{fig:spectra} would be extremely challenging, and there are a number of significant uncertainties at play, including details of how a putative new scalar couples to Standard Model (SM) particles: for instance, aside from increasing the evaporation rate and decreasing the lifetime of primordial BHs (thereby tightening current evaporation limits on their abundance), a light scalar coupled to the SM could alter the expected high-energy photon or neutrino spectra from such sources. However, without making further model-dependent assumptions, it is not possible for us to make quantitative forecasts or estimate realistic signal-to-noise ratios, so our discussion remains inevitably qualitative at this stage. With this caveat in mind, our results nonetheless highlight how theoretically well-motivated non-minimal couplings can have a significant impact on Hawking radiation spectra, well beyond what is usually assumed in minimally coupled models. This motivates follow-up work to assess whether these effects can be within observational reach.

\section{Conclusions}
\label{sec:conclusions}

It is no exaggeration to say that scalar fields play a central role across most of theoretical physics, including as test fields to probe the behaviour of gravity in the strong-field regime. A generic expectation in this context, based on considerations from quantum field theory in curved space-times, is that scalar fields should couple non-minimally to the curvature scalar $R$~\cite{Birrell:1982ix,Parker:2009uva,Fulling:1989nb}. This issue can be particularly relevant for those space-times whose Ricci scalar is non-vanishing, including nearly all non-singular (regular) BHs. Yet, the impact of such a non-minimal coupling remains to a large extent unexplored. In this pilot study, we take a first step towards systematically closing this gap: focusing on four well-motivated regular BHs taken as representative case studies (the Bardeen, Hayward, Simpson-Visser, and D'Ambrosio-Rovelli BHs), we investigate how a non-minimal scalar-curvature coupling affects their spectral properties, in particular their graybody factors and scalar Hawking emission spectra.

We demonstrate that such a non-minimal coupling can leave its imprint on BH emission spectra through a rich array of effects, ranging from an overall suppression or enhancement of the spectra, to localized features (e.g.\ bumps) in the latter. The size and direction of these effects depend not only on the magnitude and sign of the non-minimal coupling strength $\xi$, but crucially on the geometry of space-time itself, and in particular on the sign of the product between minus the $g_{tt}$ metric component in Schwarzschild-like coordinates (which we refer to as $f$) and the Ricci scalar $R$ (see Fig.~\ref{fig:fr} for the product $fR$ for the four regular BHs). The reason is that the factor $\xi fR$ controls the correction to the potential barrier induced by the non-minimal coupling, and its sign therefore determines whether the barrier is lowered or raised, changes which directly affect the transmission probability and therefore the GBFs. We find that in the conformally coupled case ($\xi=1/6$), the emission spectra change at the order unity level, except in the D'Ambrosio-Rovelli case where the spectrum is suppressed by about an order of magnitude, owing to the fact that $\xi fR$ is positive throughout the space-time, leading an overall enhancement of the potential barrier. We also considered a second case study with a large, negative non-minimal coupling ($\xi=-10^4$), inspired by Higgs inflation: in this case, we observe significant effects on all four emission spectra, but once more it is in the D'Ambrosio-Rovelli case that these effects are most significant, with an enhancement by up to five orders of magnitude, as well as the appearance of features associated to higher $l$ modes. The differences we observe in the emission spectra can be particularly important in the recently investigated scenario where light primordial regular BHs make up at least part of the dark matter in the Universe~\cite{Calza:2024fzo,Calza:2024xdh,Calza:2025mwn}. In fact, the non-minimal scalar-curvature coupling can drastically alter the evaporation history of these objects, and correspondingly loosen or tighten observational constraints on their abundance.

Overall, our results demonstrate that a theoretically well-motivated scalar-curvature non-minimal coupling can have a significant impact on the Hawking radiation spectra of regular BHs, well beyond what is usually assumed in minimally coupled models. However, as we have alluded to, these effects are likely to be extremely challenging to detect; moreover, specific model-dependent assumptions about how the scalar field couples to the Standard Model are required to be able to assess the impact, for instance, on high-energy photon or neutrino spectra from primordial regular BHs. Nevertheless, our findings motivate several interesting directions for future research. Firstly, it is important to study whether similar trends arise for other classes of regular BHs, while extending the study to rotating solutions: in this case, superradiance can play an important role, and the interplay between non-minimal couplings and superradiant instability may be non-trivial, potentially altering instability thresholds. Next, it would be interesting to study complementary signatures of non-minimal scalar-curvature couplings in the time domain, for instance within quasinormal modes and late-time tails: such a study would be particularly motivated in light of the correspondences between GBFs and quasinormal modes which have recently been brought to light~\cite{Konoplya:2024lir,Konoplya:2024vuj,Skvortsova:2024msa,Malik:2024cgb,Pedrotti:2025idg,Lutfuoglu:2025ldc,Han:2025cal,Malik:2025dxn,Lutfuoglu:2025blw,Lutfuoglu:2025pzi}. Last but not least, a more realistic assessment of whether these effects can be observed (which, as highlighted earlier, necessarily involves making assumptions on plausible scalar–matter couplings) is certainly warranted, and could further establish the observational relevance of non-minimal couplings. We leave these interesting research directions for future work.

\begin{acknowledgments}
\noindent We acknowledge support from the Istituto Nazionale di Fisica Nucleare (INFN) through the Commissione Scientifica Nazionale 4 (CSN4) Iniziativa Specifica ``Quantum Fields in Gravity, Cosmology and Black Holes'' (FLAG). M.C.\ and S.V.\ acknowledge support from the University of Trento and the Provincia Autonoma di Trento (PAT, Autonomous Province of Trento) through the UniTrento Internal Call for Research 2023 grant ``Searching for Dark Energy off the beaten track'' (DARKTRACK, grant agreement no.\ E63C22000500003). This publication is based upon work from the COST Action CA21136 ``Addressing observational tensions in cosmology with systematics and fundamental physics'' (CosmoVerse), supported by COST (European Cooperation in Science and Technology).
\end{acknowledgments}

\bibliography{nonminimally}

\end{document}